**Interpretability is in the eye of the beholder: Human versus artificial classification of image segments generated by humans versus XAI**

Romy Müller[1], Marius Thoß[1], Julian Ullrich[1,2,3], Steffen Seitz[2], & Carsten Knoll[2]

[1] *Faculty of Psychology, Chair of Engineering Psychology and Applied Cognitive Research, TUD Dresden University of Technology, Dresden, Germany*

[2] *Faculty of Electrical and Computer Engineering, Chair of Fundamentals of Electrical Engineering, TUD Dresden University of Technology, Dresden, Germany*

[3] *Faculty of Mathematics and Natural Sciences, Department of Computer Science, Machine Learning Group, Heinrich-Heine-Universität Düsseldorf, Düsseldorf, Germany*

Corresponding author:

Romy Müller
Chair of Engineering Psychology and Applied Cognitive Research
Technische Universität Dresden
Helmholtzstraße 10, 01069 Dresden, Germany
Email: romy.mueller@tu-dresden.de
Phone: +49 351 46335330
ORCID: 0000-0003-4750-7952

# Abstract

The evaluation of explainable artificial intelligence is challenging, because automated and human-centred metrics of explanation quality may diverge. To clarify their relationship, we investigated whether human and artificial image classification will benefit from the same visual explanations. In three experiments, we analysed human reaction times, errors, and subjective ratings while participants classified image segments. These segments either reflected human attention (eye movements, manual selections) or the outputs of two attribution methods explaining a ResNet (Grad-CAM, XRAI). We also had this model classify the same segments. Humans and the model largely agreed on the interpretability of attribution methods: Grad-CAM was easily interpretable for indoor scenes and landscapes, but not for objects, while the reverse pattern was observed for XRAI. Conversely, human and model performance diverged for human-generated segments. Our results caution against general statements about interpretability, as it varies with the explanation method, the explained images, and the agent interpreting them.

# 1 Introduction

A major challenge in contemporary human-technology interaction is to make black-box AI models more transparent. This can be achieved by methods of explainable artificial intelligence (XAI). In recent years, XAI has been applied in different domains such as healthcare (Srinivasu et al., 2022) and human-robot interaction (Roque & Damodaran, 2022). They can assist humans in a wide variety of tasks such as evaluating the toxicity of both mushrooms and social media (Chien et al., 2022; Leichtmann et al., 2023). A common application of XAI methods is in explaining the outputs of image classifiers based on Deep Neural Networks (DNN). For instance, during radiological diagnosis or industrial fault detection, it is desirable not only to know *that* an image classifier has detected a particular problem but also to know *why*. What areas in an image have contributed to the classifier's decision? And are these areas actually relevant, based on what areas human experts would consider when classifying the image? In particular, XAI attribution methods provide attention maps that highlight those image areas that have the highest impact on the DNN's decision[1]. Numerous XAI attribution methods have been proposed (e.g., Arras et al., 2019; Ribeiro et al., 2016; Selvaraju et al., 2017), but it is not always clear whether they generate useful visual outputs, under what conditions, and how this depends on whether you ask a human or DNN. These issues were addressed in the present study.

The motivation to conduct this study arose from a methodological gap in the literature: it is not clear how the quality of explanations should be evaluated. In particular, it is much more common for evaluations of XAI methods to solely rely on automated fidelity metrics than to take human users into account (Rong et al., 2023). Therefore, it is important to know whether the same explanations are favourable from a human perspective and from a technical perspective. A straightforward way of approaching this question empirically is to have both agents, humans and DNN, rely on the same attention maps to perform the same task: infer the image class from the image areas selected by an XAI method. To this end, the present study used a diverse set of explanations that were generated either by humans or by XAI, and on different types of images. We evaluated the explanations with human users and with a DNN, and checked in what ways these two evaluation approaches generated converging or diverging outcomes. Taken together, the article makes the following contributions:

- In two experiments with human participants, we tested how human classification performance depends on the origin of attention maps (i.e., whether these explanations reflect the implicit or explicit attention of humans, or the attention of a DNN as revealed by two XAI methods). We show that human interpretability depends on an interaction of attention map and image type. One XAI method was easily interpretable for some images, and even as interpretable as human-generated explanations, but it faced serious problems for other images. The pattern reversed for another XAI method.
- In a third experiment, we demonstrate that the findings on human classification performance generalised to human subjective ratings of the same attention maps.
- We demonstrate in what ways humans and DNN agree or disagree. This was done by comparing the results from the human experiments to DNN classification of the same attention maps. We observed a high agreement regarding XAI attention maps, but a striking disagreement in the case of human attention maps.

---

[1] The visual outputs of attribution methods are often referred to as saliency maps or attribution maps. However, both terms are either misleading or not applicable at all for visualisations of human attention. As the latter will be compared to XAI visualisations in the present study, we decided to use the term "attention maps" as it is applicable to both the image areas selected by humans and AI models.

- Taken together, we show that interpretability is in the eye of the beholder, and thus thorough evaluations of XAI methods require a concurrent consideration of three factors: the images to be classified, the explanation method, and the agent interpreting the explanation.

It is important to note that we do not propose any new DNN model or XAI method. Instead, we use standard methods that are well-known in the literature and apply them in a user study to address questions about human-computer interaction. Therefore, we neither claim nor provide proof that the DNN and XAI methods used in this study are superior to other methods. We did not intend them to be, and instead aimed to select methods that are among the most common ones. Generalising and extending the present findings to other methods will be an issue for future research.

Before specifying our research questions, we first need to consider the challenges of evaluating XAI methods in general, and the interpretability of XAI from the perspective of humans versus DNN in particular. To this end, Section 2 summarises related work in three areas: evaluation approaches for XAI methods, empirical findings about the effects of XAI attention maps on human performance, and results of previous studies comparing the interpretability of XAI attention maps to that of human-generated attention maps. This comparison also represents the core of the present study. Therefore, in Section 3, we report an experiment to analyse human classification performance of XAI versus human attention maps. In Section 4, we replicate this experiment while using a stricter method. To check whether the observed performance results extend to subjective evaluations, Section 5 reports an experiment in which participants rated the quality of the attention maps. To test how interpretability for humans compares to interpretability for DNN, Section 6 reports DNN classification performance on the same attention maps. Finally, Section 7 discusses our results and provides an outlook for future research.

## 2 Related work

### 2.1 The evaluation of explainable artificial intelligence

How can the quality of XAI methods be evaluated? Contemporary evaluations of XAI usually focus on automated metrics that describe the XAI methods' fidelity to the DNN model (Arras et al., 2022; Vilone & Longo, 2021). This comes with two problems. The first one is methodological: evaluation metrics often suffer from insufficient reliability and validity (Arras et al., 2022; Tomsett et al., 2020; Wang et al., 2020). The second problem is more conceptual: fidelity-based evaluation metrics only assess whether an XAI method has actually highlighted the areas that were most important to the DNN, but not whether these areas are meaningful from the perspective of human decision-makers. This is a serious shortcoming, because XAI attention maps that do not make sense to humans can cause distrust in the DNN, even when its classification performance is high (Nourani et al., 2019). Therefore, human interpretability should complement measures of fidelity in XAI evaluations (Rong et al., 2023).

A common but rather indirect way of assessing human interpretability is to gather subjective ratings. This can be done by asking humans to judge the quality of XAI outputs (Sundararajan et al., 2019), to indicate how confident they are that the AI will make correct decisions (Karran et al., 2022), or to reveal how willing they are to use the AI in the future (Ebermann et al., 2023). However, this subjective approach presupposes that humans are aware of their information needs, and able to judge the ability of XAI outputs to fulfil these needs. Therefore, a promising alternative way to assess human interpretability is to directly measure whether human information needs are fulfilled. This can be achieved by testing whether XAI attention maps enhance human task performance.

## 2.2 Assessing interpretability via human task performance

Does XAI support human task performance? Contrary to the prevailing optimism about XAI attention maps among the public, recent studies suggest that their usefulness is quite limited. They may help users detect bugs and biases in AI models (Balayn et al., 2022; Colin et al., 2022; Ribeiro et al., 2016), but do they also enhance the performance of human-AI teams in the actual task they intend to explain? That is, do they allow for more accurate image classification? Two recent studies cast doubt on the usefulness of XAI attention maps for image classification (Chu et al., 2020; Nguyen et al., 2021). In these studies, human classification performance with explanations was not superior to performance without them. Moreover, the lack of XAI benefits did not depend on XAI quality: human performance appeared to be unaffected by the attention maps, regardless of whether they selected highly relevant or completely irrelevant image areas. However, it needs to be considered that classifying a fully visible image may not be the most suitable task for assessing the benefits of XAI attention maps. Such benefits may be highly task-specific. For instance, in one study, XAI attention maps supported users in detecting the biases of a DNN and in inferring its visual strategies, but not in understanding failure cases (Colin et al., 2022). Thus, it is important to use evaluation paradigms in which there is a close match between XAI capabilities and human information requirements.

One such paradigm is the classification of image segments generated via XAI attributions (Knapič et al., 2021; Lu et al., 2021; Rong et al., 2021; Slack et al., 2021). For this purpose, gradual attention maps are transformed into binary masks, which can be overlaid on the image. There are two versions of this basic approach. First, only the most important areas can be kept visible, while the rest of the image is hidden. If the areas are useful to humans, accurate classification should still be possible, even though only a small part of the image remains visible. A second version of the paradigm reverses this principle, hiding exactly those image areas that are deemed most relevant by the XAI. If the areas are useful to humans, classification performance should deteriorate.

Using this segment classification paradigm, a recent study not only investigated whether XAI was beneficial to human performance, but how these performance effects varied between four XAI methods (Lu et al., 2021). The authors found clear evidence that the XAI methods had different impacts on human performance. For instance, a method that produced connected attributions enabled more accurate classification than methods that produced more pixelated attributions. At the same time, these human-related results did not match those obtained with automated fidelity metrics. Such discrepancies highlight the importance of involving humans in the evaluation of XAI. Apparently, user studies can generate insights beyond those obtained with automated metrics.

So far, we have argued that human interpretability, particularly when assessed via objective performance measures in suitable paradigms, can be a valuable criterion for evaluating XAI methods. However, it is not desirable either to favour XAI methods that make sense to humans but provide inaccurate information about the inner workings of a DNN. Therefore, it is important to know whether an XAI method's interpretability and its fidelity go hand in hand. Do image areas that make sense to humans also make sense to DNN? Such correspondence in information needs can be investigated by comparing human and DNN classification performance on image segments that have been generated by either XAI methods or other humans (gathered via eye tracking or manual selection). Some previous studies have investigated how effectively humans can classify XAI versus human attention maps, while others have investigated the same for DNN. In the following section, we will review these studies, explain why they leave important questions unanswered, and describe our own contribution to closing this gap.

**2.3 The interpretability of human versus DNN attention**

Are attention maps that reflect human visual processing more interpretable than XAI that reflects DNN processing? Several previous studies have empirically investigated whether humans and DNN attend to similar image areas (e.g., Das et al., 2017; Ebrahimpour et al., 2019; Hwu et al., 2021; Lanfredi et al., 2021; Rong et al., 2021; Singh et al., 2020; van Dyck et al., 2021; Zhang et al., 2019). Usually, these studies found low to medium similarity, and we have provided a detailed review of their findings in a recent article (Müller, Dürschmidt, et al., 2023). But what does this lack of similarity mean? In principle, it could mean at least three things: that a DNN does not use suitable image areas for classification, that XAI methods do not adequately reflect which areas were actually relevant to the DNN, or that humans are mistaken in terms of which areas are most relevant. This ambiguity cannot be resolved by only assessing the similarity between human and XAI attention maps. Instead, these attention maps must actually be used for classification by humans and DNN. The resulting classification performance can contribute to answering two questions: whether XAI attention maps are more or less interpretable than human attention maps, and how this depends on whether it is a human or DNN who needs to interpret them. Two previous studies suggest divergent hypotheses. On the one hand, it has been reported that humans can better classify images based on the areas selected by XAI than by other humans (Zhang et al., 2019). On the other hand, it has been reported that DNN can better classify images based on the areas selected by humans than by XAI (Liu et al., 2023; Rong et al., 2021). Does this mean that humans can better interpret XAI, while DNN can better interpret humans? Probably not, as several methodological differences between the two studies complicate their integration.

In the study by Zhang et al. (2019), human attention maps were generated by a first group of participants who saw images of complex scenes split into 50 segments, and had to manually order these segments according to their relevance for classification. It was found that participants assigned the most relevance to the class-defining object, so their attention maps lacked the rich information provided by the scene context. Conversely, DNN assigned high relevance to this context, and thus more context information was included in XAI attention maps. As scene context is of paramount importance for human scene classification (Oliva, 2005; Torralba & Oliva, 2003), human-generated segments were less interpretable for other humans than XAI-generated segments. Accordingly, when a new group of participants saw the segments successively in their order of relevance assigned by either humans or DNN, they needed fewer XAI segments than human segments to infer the correct image class. From a practical perspective, this might call into question the use of humans as a "ground truth" for relevance estimation. However, before drawing such far-reaching conclusions, we need to critically consider the methodological approach of the study. Specifically, human attention maps were based on manual ordering of pre-defined image segments. It is not too surprising that when asked to select the most relevant segments to classify an object in a scene, humans first select all segments pertaining to this object. This may not correspond to the areas they actually use for classification. Thus, it is important to contrast different procedures for eliciting human attention maps. The results might change when human image segments are generated in a way that is closer to human attentional processes, for instance by tracking eye movements during classification.

The suitability of eye movements for eliciting human attention maps has been corroborated by other studies. Rong et al. (2021) recorded eye movements while participants performed fine-grained classification of close-up images showing one of two highly similar bird species. The authors found that the attention maps derived from these eye movements were more discriminative than XAI attention maps – they more specifically targeted the feature that actually differentiated between the two bird species. Accordingly, when a DNN had to classify image segments generated by either humans or XAI, a smaller image area was sufficient to reach high performance with human segments. Similarly, other studies also suggest that human attention can improve the performance of DNN (Boyd et al., 2023;

Karargyris et al., 2021; Lai et al., 2020). A particularly interesting, recent study has shown that the fidelity of human attention can be higher than that of XAI attributions (Liu et al., 2023). The authors compared attention maps that were either based on human eye movements or based on four XAI methods. Surprisingly, eye movements had a much higher fidelity to the DNN, although obviously, humans were not actually explaining the inner workings of the DNN. Apparently, the computations that actually took place in the model were well-aligned with human vision, even though XAI could not adequately explain this.

All these previous findings on the interpretability of attention maps are hard to integrate, because the studies differed with regard to several methodological factors. They used different types of human attention maps (i.e., manual selection vs. eye tracking) and XAI attention maps (i.e., SHAP vs. Grad-CAM, CAM, IG, and IxG) on different types of images (i.e., complex scenes vs. individual objects), and the attention maps had to be interpreted by different agents (i.e., humans vs. DNN). To understand the impact of these methodological variations, we manipulated them in one and the same study.

## 2.4 Present study

In the present study, human and XAI attention maps were transformed into binary image segments that only revealed the most important areas of an image. Their interpretability was evaluated by assessing how effectively they supported classification. We asked whether this interpretability depended on who has generated the segments, what type of image they stemmed from, and who had to interpret them. First, to investigate how interpretability depended on who has generated the image segments, we compared four segment types, two of them generated by humans (i.e., gaze, drawing) and two generated by XAI (i.e., Grad-CAM, XRAI). Concerning the human segment types, gaze reflects which areas humans actually looked at during classification, whereas drawings reflects which areas humans explicitly considered most relevant. These two elicitation procedures come with different cost and benefits (for a detailed discussion see Müller, Dürschmidt, et al., 2023). While gaze avoids the problem of humans being unaware of their true information needs (Zhang et al., 2019), it is prone to systematic viewing biases, such as central fixation bias (Tatler, 2007) or attentional capture by salient but task-irrelevant features (Itti & Koch, 2000). Moreover, gaze-based attention maps were found to be less similar to XAI than drawing-based attention maps (Müller, Dürschmidt, et al., 2023). To generate XAI segments, we used Grad-CAM (Selvaraju et al., 2017) and XRAI (Kapishnikov et al., 2019), which have both been deemed to be particularly human-friendly, because they provide connected regions instead of isolated pixels. Despite this similarity, it has been shown that the attention maps generated by Grad-CAM are more similar to human attention than those generated by XRAI, at least for Convolutional Neural Networks (Morrison et al., 2023). Thus, it will be interesting to investigate whether these two XAI methods will also differ in their effects on human performance. Comparing two different XAI methods was necessary to both generalise and differentiate our findings. Regarding generalisation, it is important to ensure that our results are not merely an artefact of the capabilities and limitations of a particular XAI method. Regarding differentiation, we would like to understand whether different XAI methods are more or less suitable in different situations. For instance, one XAI method might be more interpretable when classification relies on specific objects (e.g., lighthouses), but another one when classification relies on global scene properties (e.g., deserts).

This directly leads to the second influence on interpretability that was investigated in the present study, namely the differentiation between image types. Our segments stemmed from images of either objects, indoor scenes, or landscapes. The rationale for choosing these image types was that the areas needed for their classification varied in specificity or diversity. That is, while a classification of objects mainly relies on a particular image area, the classification of indoor scenes largely draws on object-to-object relations and the classification of landscapes exploits global scene properties (Greene & Oliva, 2009; Henderson, 2017; Torralba & Oliva, 2003; Võ et al., 2019). Accordingly, these image types lead to remarkably different attention maps, and different degrees of similarity between humans and XAI (Müller, Dürschmidt, et al., 2023). For objects, this similarity was highest, but also most dependent on the procedure of eliciting human attention maps (i.e., XAI was much more similar to segments generated by drawing than gaze). In contrast, for landscapes human-XAI similarity was lowest, independent of the elicitation procedure. Indoor scenes were located somewhere in between. However, what remains unclear from these findings is whether the differences in similarity go along with differences in interpretability.

The third influence on interpretability that was investigated in the present study is whether image segments had to be classified by humans or DNN. This is important, because previous studies suggest different outcomes, either reporting that humans benefit more from XAI or that DNN benefit more from human attention. However, it is unclear whether this double dissociation will hold when humans and DNN classify the same segments. To test how easily humans can interpret image segments, we conducted three experiments. In Experiment 1, the basic task was to indicate whether an image segment matched a previously shown class label. For instance, participants saw the word "desert" followed by an image segment, and had to press one key if they thought the segment indeed was part of a desert, and another key if they did not. In half of the trials, the label and segment were compatible (i.e., same class, e.g., desert), whereas in the other half they were incompatible (i.e., the label referred to one of the remaining classes, e.g., lighthouse, windmill, office, dining room, or wheat field). Experiment 2 served to replicate and extend our findings by making classification more challenging and specific. To this end, incompatible trials only used labels of the same image type as the segment (e.g., desert segments were preceded by the word wheat field that also refers to a landscape, but not by a word that refers to an object or indoor scene). Experiment 3 tested how closely the classification results matched with subjective interpretability by asking participants to rate how well each image segment depicted its respective class. Finally, to assess how interpretability depended on the agent interpreting the segments, we fed all human and XAI segments into the DNN that was supposed to be explained by the XAI segments, and compared DNN classification performance to human classification performance. Ultimately, this should allow us to examine whether the human interpretability of XAI segments goes hand in hand with its fidelity to the DNN model, or under what conditions these two criteria for evaluating XAI methods diverge.

## 3 Experiment 1

In the first experiment, participants had to decide whether image segments matched a previously shown class label, and non-matching labels were randomly drawn from all alternative classes. We hypothesised that humans can classify human segments better than XAI segments. In this regard, we diverged from Zhang et al. (2019), who reported inferior performance with human segments. This divergence resulted from the fact that their presumed reason for human inferiority (i.e., loss of scene context) does not apply in the present study. As our image segments were derived from a previous study (Müller, Dürschmidt, et al., 2023), we knew that both procedures for eliciting human image segments (i.e., gaze and drawing) resulted in an inclusion of ample scene context. Therefore, we went with the reasoning put forward by Rong et al. (2021) and expected superior classification performance with human segments, assuming that these segments are more focused on diagnostic features than XAI segments. Moreover, we expected human classification performance to be indifferent to the specific type of human segment (i.e., gaze or drawing). First, we knew that these segments largely overlapped in their contents, despite differences in visualisation details (Müller, Dürschmidt, et al., 2023). Second, although human classification of visual materials can benefit from seeing another person's gaze (for a review see Emhardt et al., 2023), the usefulness of transferring eye movements typically does not differ from that of transferring mouse movements (Müller et al., 2014; Müller et al., 2013). Thus, we expected human classification performance to be similar for the two human segment types and higher for human segments than XAI segments.

Moreover, we hypothesised that the effects of segment type would interact with image type. For objects, we expected no major differences between the four segment types. This is because object segments were quite similar across segment types, typically highlighting the relevant object (Müller, Dürschmidt, et al., 2023). However, informal observations suggested that Grad-CAM was slightly inferior to XRAI (e.g., tending to highlight the lower part of lighthouses and windmills, or highlighting distractor objects). Thus, for object images we expected performance to be somewhat lower with Grad-CAM than the other segment types. For indoor scenes, we expected large differences between segment types. For one, segments might be harder to tell apart between different indoor scene classes (e.g., tables are present in both offices and dining rooms). This should amplify the importance of selection specificity, which is higher for humans than XAI (Rong et al., 2021; Zhang et al., 2019). Moreover, indoor scenes are challenging to classify for computational models (Quattoni & Torralba, 2009). Informal inspections of our segments revealed that XRAI produced some puzzling results (e.g., highlighting a window to explain dining room). Thus, we expected performance for indoor scenes to be higher with human segments than XAI segments. Moreover, again we expected no major differences between the two human segment types. If anything, performance might be somewhat lower with gaze segments, because eye movements easily get distracted by salient but irrelevant features, which could result in more fragmented segments. Finally, for landscapes we expected no differences between segment types. Although their areas had a lower overlap than for objects and indoor scenes (Müller, Dürschmidt, et al., 2023), these differences presumably are non-consequential. This is because large-scale outdoor scenes are characterised by global properties, with major parts of the image being equally informative (Greene & Oliva, 2009; Torralba & Oliva, 2003). Thus, we expected all segment types to provide information of similar relevance for classification despite differences in their selected areas.

### 3.1 Method

#### 3.1.1 Open Science

All images and human participant data are made available via the Open Science Framework (https://osf.io/pvmfj/). The source code for our DNN, XAI, and attention maps is made available on GitHub (https://github.com/cknoll/Humans-vs.-CNN-Effects-of-task-and-image-type).

#### 3.1.2 Participants

Twenty-four participants (corresponding to all possible block orders of the four segment types) were recruited via the TUD Dresden University of Technology participant pool (*ORSEE*, Greiner, 2015), 13 of them were female and 11 were male, and their age ranged from 19 to 73 years ($M = 27.8$, $SD = 14.1$). They took part in the experiment in exchange for 10 € or partial course credit. None of them had participated in the experiment from which the image segments were generated (Müller, Dürschmidt, et al., 2023). Moreover, participants had to be native speakers of German. The research was approved by the Ethics Committee at the TUD Dresden University of Technology (file sign: SR-EK-400092020), participants provided written informed consent, and all procedures followed the principles of the Declaration of Helsinki.

*3.1.3 Apparatus and stimuli*

*Lab setup*. All experimental sessions took place in a dimly lit lab room at TUD. Stimuli were presented on a 24" LCD display with a resolution of 1920 x 1080 pixels at a refresh rate of 60 Hz. Responses were entered with a standard QWERTZ keyboard and computer mouse. The experiment was programmed with the SR Research Experiment Builder (SR Research Ltd., Ontario, Canada).

*Types of screens*. During the experiment, six types of screens were shown: label screens and segment screens, as well as additional screens for instructions, ratings, feedback, and demographics. Label screens presented the verbal label for one class (i.e., lighthouse, windmill, office, dining room, desert, or wheat field). This word was presented centrally on a white background in black font (Tahoma, 30 pt). The labels and all other verbal materials were presented in German. Segment screens presented one image per screen at a resolution of 1024 x 1024 pixels on a white background. Only the most important 5 % of the image were visible (see below for a description of the segment generation procedure) and the rest was hidden (i.e., covered by a black mask). Rating screens required participants to indicate how many trials they thought they had solved correctly by selecting a number between 0 and 100 %. Feedback screens presented the words "richtig" or "falsch" (German for correct and incorrect) in green and red font, respectively, on a white background (Tahoma, 30 pt). Instruction screens explained the task, and demographics screens asked participants to input their age and gender.

*Image types*. Example stimuli are presented in Figure 1 and the full set of stimuli is available at the Open Science Framework. We presented 240 image segments in total, visualising the most relevant areas of 60 images according to four segment elicitation procedures that created four segment types (i.e., gaze, drawing, Grad-CAM, XRAI). Moreover, we used 20 additional, fully visible images (i.e., without segment extraction) for practice.

The original images that were used for segment generation were taken from the Places365 dataset (Zhou et al., 2017). This dataset offers about 1.8 million images from 365 different classes with up to 5,000 images per class. We chose this specific dataset due to its wide variety of complex natural images from different scene types (i.e., objects, indoor scenes, landscapes). This distinguishes the dataset from more object-focused datasets like ImageNet (Deng et al., 2009). In the present study, the complexity of natural scenes was important as it allowed us to distinguish between images that rely on singular, diagnostic objects, systematic object-to-object-relations, and global scene properties. Moreover, Places365 provides images at a much higher and more consistent spatial resolution than other datasets like ImageNet. In this way, small image segments can easily be recognised, even when they only span 5 % of the entire image area.

The 60 images used in the main experiment consisted of 20 images per image type (i.e., objects, indoor scenes, landscapes). Moreover, each image type consisted of two classes, resulting in 10 images per class. The two classes within an image type were chosen to be highly similar, in order to make classification more challenging. For the image type *objects*, scenes included a clearly discernible, localised object that defined the class (i.e., lighthouse, windmill). These objects were embedded in a large-scale scene context, which could be more or less typical (e.g., lighthouses in front of a coastline vs. in an urban area). Images of *indoor scenes* showed a room with a specific function (i.e., office, dining room), and all of them included chairs and tables. Images of *landscapes* presented a wide, open outdoor scene (i.e., desert, wheat field) that consisted of large, uniformly structured and coloured areas, and some landscapes contained additional objects (e.g., agricultural machinery, houses).

**Figure 1**

*Stimulus examples for each combination of segment type and image type.*

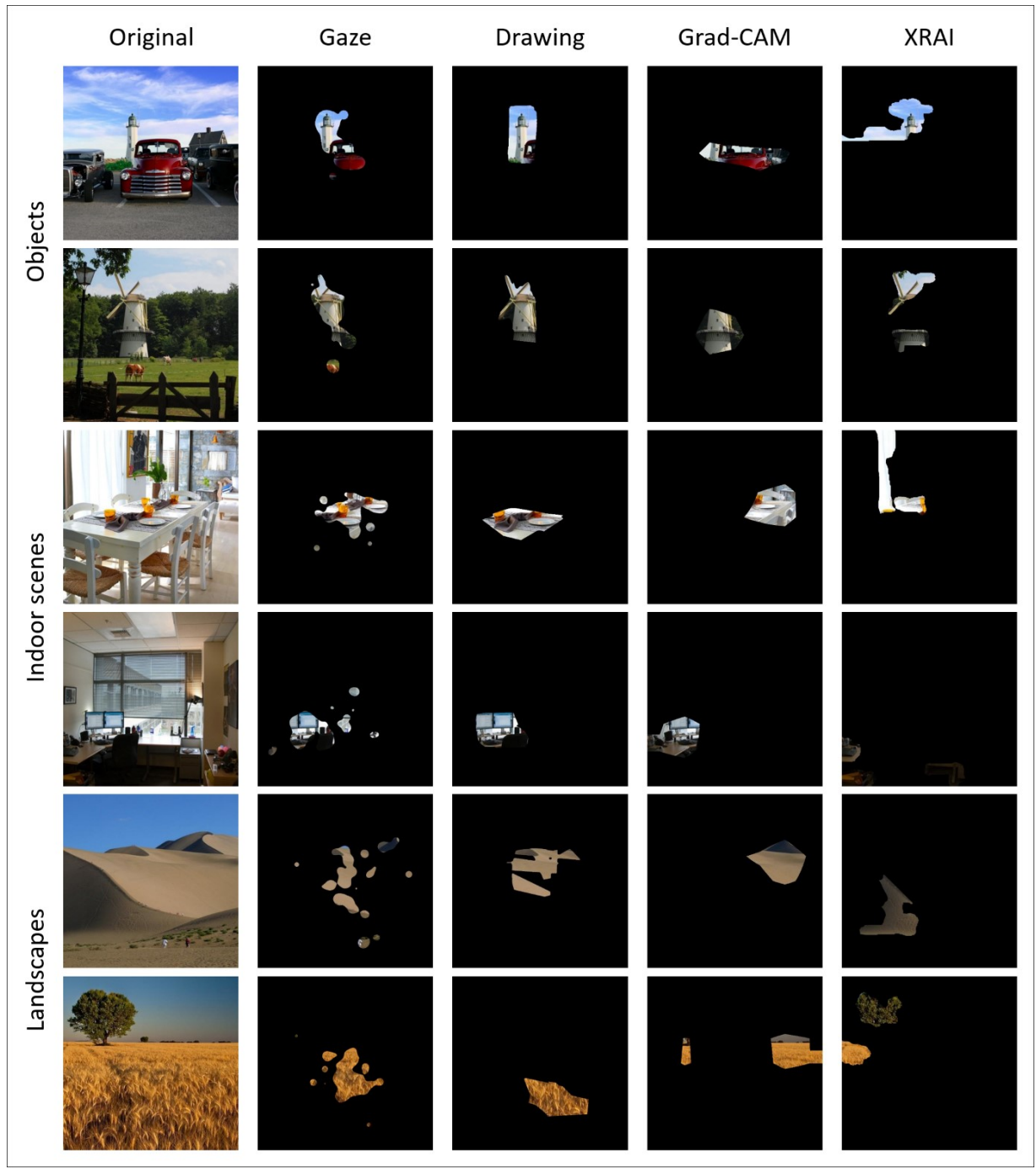

*Image segment generation.* These procedures for generating the four segment types selected the most relevant 5 % of each image (see Figure 2). Segments could consist of one or several parts and the value of 5 % referred to the sum of all parts. The rest of the image was covered with a black mask. A segment size of 5 % was chosen because it adequately differentiates between human and XAI segments, while larger segments can easily render human and XAI too similar (Rong et al., 2021). Moreover, segment sizes above 5 % could not be implemented for several of our object images, because in the human conditions (i.e., gaze, drawing) large parts of the image never received more attention even when summed over all participants who contributed to segment generation.

The human segments were based on an earlier study with 25 participants and the full details of segment generation are provided in the respective article (Müller, Dürschmidt, et al., 2023). For generating *gaze segments*, participants' eye movements were tracked while they classified the fully visible image by pressing one of six keys. They were asked to intentionally look at the relevant image areas (i.e., gaze-pointing). Based on these eye movement recordings, a Gaussian kernel was applied in an area of 2 degrees of visual angle (58 pixels) around each fixation, so that the weight decreased (proportional to a symmetric Gaussian 2D distribution) from the centre to the outside. This kernel was multiplicatively weighted by fixation duration, so that longer fixations had more impact. The weighted kernels of all fixations were added over all participants. Subsequently, binary segments were created by keeping only those 5 % of the image area visible that received the most weight.

To generate *drawing segments*, participants had to draw a polygon around the image area they considered most relevant for classification, clicking the intended location of a polygon corner with their mouse. All image areas inside a participant's polygon received a value of 1 and all areas outside received a value of 0. After summing the polygons of all participants, we again only kept the most highly weighted area visible. To this end, a gradualisation procedure was applied to make sure that exactly 5 % were selected.

Our XAI segments were based on a ResNet-152 architecture (He et al., 2016) and classification decisions of this DNN were explained by Grad-CAM (Selvaraju et al., 2017) and XRAI (Kapishnikov et al., 2019). We chose ResNet and Grad-CAM, because they are by far the most frequently used DNN model and XAI method in previous user studies investigating the effects of XAI on human performance. Using these standard methods will make our results more comparable to the existing literature. We additionally chose XRAI, because it has been argued that this segmentation-based XAI algorithm can mitigate problems of Grad-CAM from a user perspective (Kapishnikov et al., 2019). Given that the DNN and XAI methods were not developed by us but are very common technologies, we will only provide a brief overview of them here. For more details concerning the technical implementation, we refer the reader to the respective original publications.

In short, the ResNet-152 consisted of many convolutional layers to extract features, followed by a fully connected layer (i.e., classification head), which was used to calculate a score for each class. The distinct feature of the ResNet architecture is that it has a special internal structure including residual connections, which are leapfrogging some of the convolutional layers and thus providing a secondary data path. This significantly reduces the problem of vanishing gradients during training, which otherwise would limit the number of layers. The ResNet-152 was implemented using the PyTorch machine learning framework. It was trained from scratch on the Places365 dataset for 10 epochs on a GPU cluster of the Center for High Performance Computing (ZIH) at the TUD Dresden University of Technology. We followed the standard training procedure proposed in the original paper. This procedure includes resizing the images and extracting 224 x 224 random crops, which are then used in the training process.

Grad-CAM (Selvaraju et al., 2017) rests on the assumption that the activation values of feature maps in the final convolutional layer of the DNN represent the location of those features in the input image. Thus, the algorithm weights these activations based on their relative contribution to the class decision and combines them into an importance map. This map can be upsampled to the original image size. From the upsampled map, we extracted the 5 % of all pixels with the highest importance values.

XRAI (Kapishnikov et al., 2019) is based on an XAI method called "Integrated Gradients" (Sundararajan et al., 2017) combined with a particular segmentation procedure (Felzenszwalb & Huttenlocher, 2004). Integrated Gradients provides importance values for each pixel. This is done by iteratively applying the DNN to variations of the original image blended with a black or white image baseline and accumulating (i.e., integrating) the derivative (i.e., gradient) of the output value with respect to each input pixel. While the resulting pixel map shows the impact of individual pixels on DNN classification, it is poorly interpretable for humans due to its sparsity and fragmentation. To increase interpretability, the XRAI method uses the Felzenszwalb algorithm for decomposing the original image into segments. These segments are then weighted with the attributions generated by Integrated Gradients to create a dense importance map for the image. To ensure comparabilty with the other segment types, we smoothened the XRAI importance map by averaging each pixel value of the map over its direct neighbourhood. After this gradualisation procedure, we selected the 5 % of the image with the highest importance.

**Figure 2**

*Generating image segments that uncover the most important 5 % of the image area, depending on segment type.*

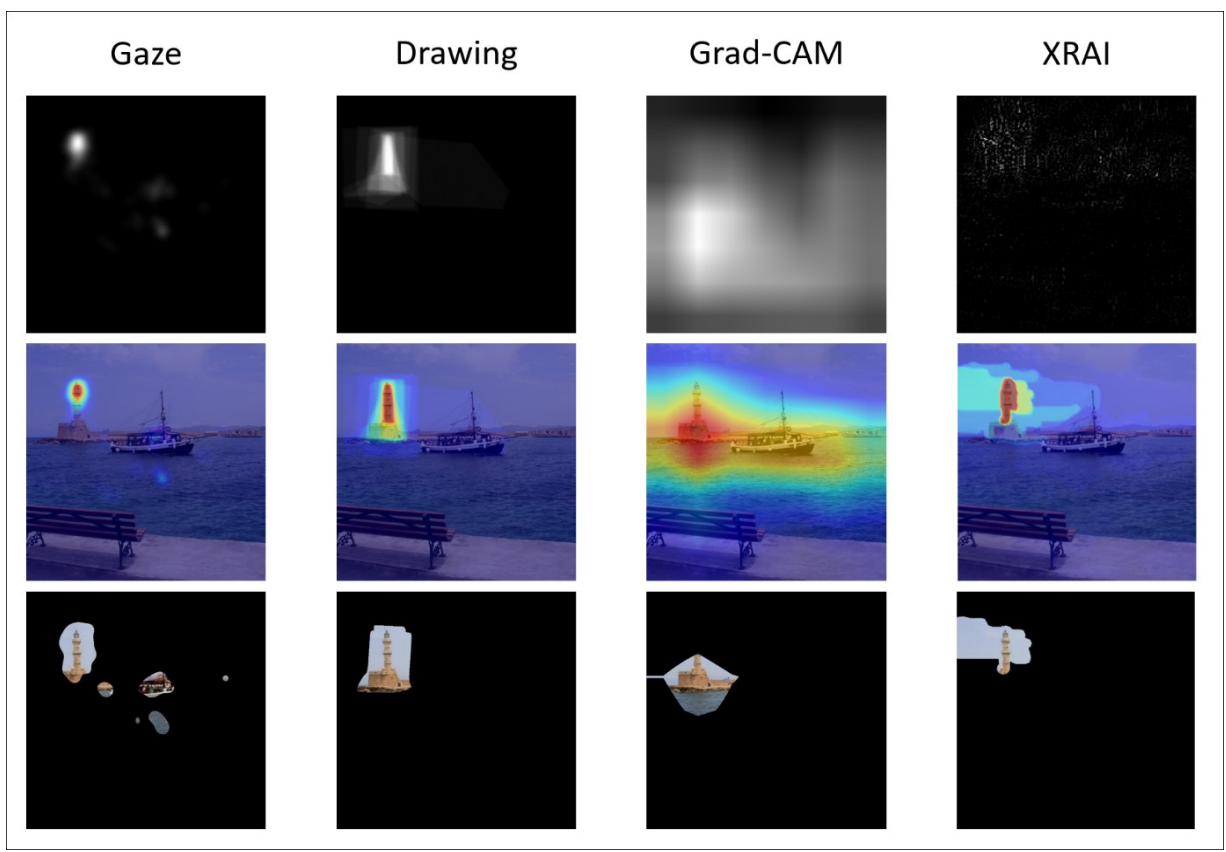

### 3.1.4 Procedure

*Design*. An overview of the procedure is provided in Figure 3. Throughout the experiment, participants had to decide whether an image segment belonged to the class indicated by a label beforehand. In a within-participants design, we varied two factors: segment type (gaze, drawing, Grad-CAM, XRAI) and image type (objects, indoor scenes, landscapes). Segment type was varied between blocks and image type was varied between trials.

*Overview of the experiment*. The experiment took between 20 and 30 minutes. A session started with participants receiving a brief summary of the procedure and providing written informed consent.

Specific task instructions were provided on-screen. Importantly, although these instructions stated that the experiment dealt with XAI, participants were not told which segments were generated by XAI versus humans or even that some segments were not generated by XAI. After the instruction, participants performed a practice block (20 trials) in which they did not see segments but fully visible original images (i.e., without masks) from all six classes. None of these images reappeared in the main experiment. Participants' task was the same as in the main experiment (see below), but they additionally received feedback on the correctness of their responses.

*Block structure of the experiment*. The main experiment consisted of four blocks with 60 trials each (i.e., 60 images, presented in random order). The blocks corresponded to the four segment types (i.e., gaze, drawing, Grad-CAM, XRAI) and their order was counterbalanced across participants. After each block, participants were asked to rate how many trials they thought they had solved correctly on a scale between 0 and 100 %.

*Procedure of a trial*. The basic procedure of a trial was identical for each block. A trial started with the presentation of a label (corresponding to one of the six classes, e.g., "desert"). This label remained on the screen for 1000 ms and was then followed by a segment that remained on the screen until participants submitted their keypress response. They had to indicate as quickly and accurately as possible whether the image matched the label by pressing the X key (match) or N key (mismatch). Participants were asked to rest their index fingers on these keys throughout the experiment. In the main experiment, they no longer received feedback about the correctness of their responses.

**Figure 3**

*Overview of the procedure of Experiments 1 and 2. CR = correctness rating.*

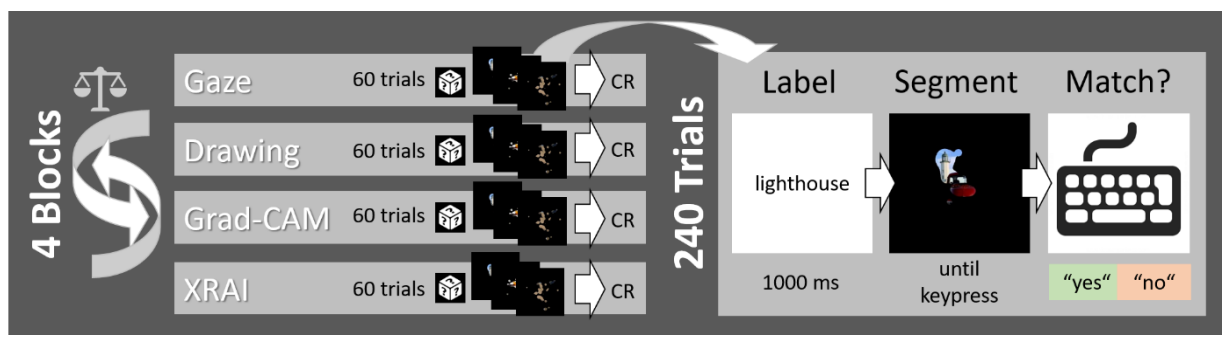

### 3.1.5 Data analysis

To quantify human performance, we statistically analysed mean response times and error rates using 4 (*segment type: gaze, drawing, Grad-CAM, XRAI*) x 3 (*image type: objects, indoor scenes, landscapes*) repeated measures ANOVAs. An alpha value of *p* = .05 was used to determine statistical significance, and all pairwise comparisons were performed with Bonferroni correction. If the sphericity assumption was violated, a Greenhouse-Geisser correction was applied and the degrees of freedom were adjusted.

### 3.2 Results

#### 3.2.1 Response times

Response time was defined as the latency from segment onset until participants pressed a key to indicate whether the label and segment matched. We excluded all trials with response times higher than 3000 ms (2.0 % of the data). The ANOVA revealed a main effect of segment type, $F(3,69) = 23.372$, $p < .001$, $\eta p^2 = .504$. This effect was due to participants responding more slowly to XRAI segments (1104 ms) than to gaze, drawing, and Grad-CAM segments (956, 936, and 968 ms, respectively), all *p*s < .001, while responses did not differ between these three segment types, all $ps > .9$. Moreover, a main effect of image type, $F(2,46) = 27.492$, $p < .001$, $\eta p^2 = .544$, resulted from faster responses for objects (918 ms) than indoor scenes and landscapes (1028 and 1027 ms, respectively), both $ps < .001$, while response times to these two scene-centric image types did not differ, $p > .9$. The two main effects were qualified by a significant interaction, $F(4.1,93.5) = 12.566$, $p < .001$, $\eta p^2 = .353$, indicating that the direction and magnitude of the differences between segment types depended on image type (see Figure 4A). For objects, both human segment types yielded faster responses than both XAI segment types. That is, responses were faster for gaze and drawing segments (881 and 850 ms, respectively) than for Grad-CAM and XRAI segments (980 and 962 ms, respectively), all $ps < .012$. Neither the two human segment types differed from each other nor the two XAI segment types, both $ps > .9$. For indoor scenes, only responses to XRAI segments (1210 ms) were slower than responses to gaze, drawing, and Grad-CAM segments (980, 985, and 938 ms, respectively), all $ps < .001$, while the latter three segment types did not differ from each other, all $ps > .4$. The same pattern was observed for landscapes: only responses to XRAI segments (1141 ms) were slower than responses to gaze, drawing, and Grad-CAM segments (1006, 974, and 987 ms, respectively), all $ps < .001$, while the latter three segment types did not differ from each other, all $ps > .9$.

**Figure 4**

*Response times (A) and error rates (B) for Experiment 1, depending on segment type and image type. Error bars represent standard errors of the mean.*

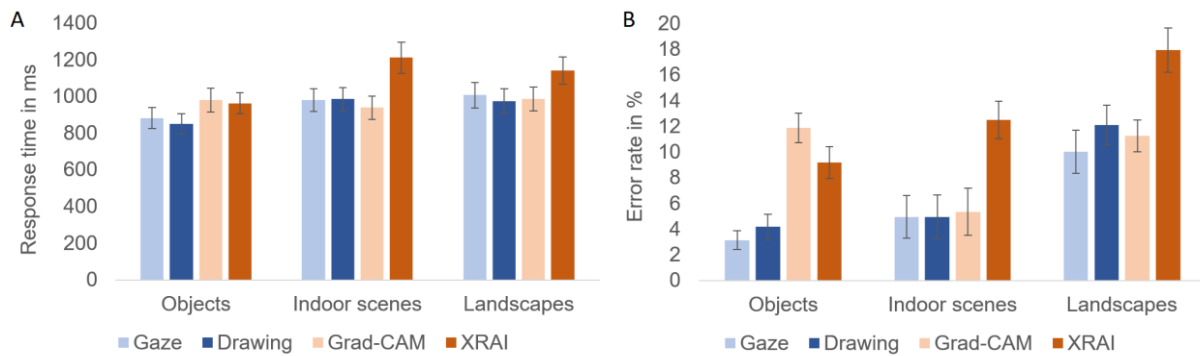

### 3.2.2 Error rates

Errors were defined as participants indicating that a label and segment matched when in fact they did not, or vice versa. There was a main effect of segment type, $F(3,69) = 17.545$, $p < .001$, $\eta p^2 = .433$. This effect was due to participants committing more errors with XRAI segments (13.2 %) than gaze and drawing segments (6.0 and 7.1 %, respectively), both $ps < .001$. For Grad-CAM segments (9.5 %), the results were less clear: while they produced more errors than gaze segments, $p = .014$, they neither were significantly worse than drawing segments, $p = .084$, nor significantly better than XRAI segments, $p = .055$[2]. Moreover, the two human segment types did not differ from each other, $p > .9$. Second, a main effect of image type, $F(2,46) = 14.483$, $p < .001$, $\eta p^2 = .386$, resulted from more errors for landscapes (12.8 %) than objects and indoor scenes (7.1 and 6.9 %, respectively), both $ps < .001$, while objects and indoor scenes did not differ, $p > .9$. Third, the main effects were qualified by a significant interaction, $F(3.9,90.6) = 5.437$, $p < .001$, $\eta p^2 = .191$, indicating that the differences between segment types depended on image type (see Figure 4B). For objects, both human segment types yielded more correct responses than both XAI segment types. That is, error rates were lower for gaze and drawing segments (3.1 and 4.2 %, respectively) than for Grad-CAM and XRAI segments (11.9 and 9.2 %, respectively), all $ps < .011$. Neither the human segment types differed from each other nor the XAI segment types, both $ps > .2$. For indoor scenes, only responses to XRAI segments (12.5 %) were less correct than responses to gaze, drawing, and Grad-CAM segments (4.9, 4.9, and 5.4 %, respectively), all $ps < .018$, while the latter three segment types did not differ, all $ps > .9$. For landscapes, gaze and Grad-CAM segments (10.0 and 11.3 %, respectively) yielded fewer errors than XRAI segments (17.9 %), while drawing segments (12.1 %) did not significantly differ from any other segment type, all $ps > .05$.

---

[2] Both comparisons passed the significance level when excluding a single participant with extreme error rates for indoor scenes. For this image type, the participant committed between 40 and 45 % errors for gaze, drawing, and Grad-CAM segments, while the mean of all other participants was between 3.5 and 3.7 %, and not a single other participant committed more errors than 15 %. However, we saw no theoretical reason for excluding this participant, and thus decided to keep him in the sample – also because we think that one should be sceptical about significant effects that are highly dependent on individual participants. If the non-significant trends are meaningful, they should become significant in a replication (Experiment 2).

### 3.3 Discussion

Experiment 1 investigated whether humans could better interpret image segments generated by other humans (i.e., gaze, drawing) or by XAI (i.e., Grad-CAM, XRAI). Overall, participants were able to quickly and accurately classify human segments regardless of whether they had been generated from eye movement recordings or manual selections. Moreover, participants' ability to classify XAI segments depended on the specific XAI method. XRAI consistently slowed down responses and increased error rates compared to the human segments. In contrast, Grad-CAM only deteriorated performance for objects, but it was just as easy to use as human segments for indoor scenes and landscapes.

Why were XRAI segments so hard to classify for human observers? An inspection of these segments provided several hints. First, for indoor scenes XRAI preferably selected unitary areas with clear edges, presumably due to its segmentation approach to generating XAI highlights. Thus, it selected areas like the seats of chairs, legs of tables, windows, or computer screens. Apparently, this selection was not particularly distinctive. In contrast, Grad-CAM tended to select the dishes on dining room tables, which more clearly gave away the identity of the room. Similarly, XRAI often selected computer monitors, but when they appeared as dark squares in the small segment area without sufficient context, they were hard to identify. In line with this apparent object focus, XRAI also tended to select distractor objects like trees in wheat fields. Conversely, Grad-CAM included more diverse context instead of focusing on one particular object. The very same might contribute to explaining the difficulties participants experienced with Grad-CAM for objects. Some Grad-CAM segments did not seem well-positioned but mostly targeted the lower part of a lighthouse or windmill, including its scene context, while the more diagnostic upper part remained hidden. A similar criticism of Grad-CAM has been put forward before, namely that it does not precisely target relevant objects, while XRAI supposedly avoids this problem (Kapishnikov et al., 2019). The problems of XRAI may not have come to light in that study, because it was only evaluated on the highly object-centric ImageNet dataset (Deng et al., 2009).

Before drawing conclusions about the interpretability of human versus XAI segments, we need to address a methodological problem. In principle, it was possible for participants to work by exclusion instead of actually identifying a segment. Even when not knowing for sure that a dark square was a computer monitor in an office, they could exclude that it stemmed from a desert or lighthouse. This is because low-level physical properties of scene patches (e.g., structure and colour) reliably tell apart superordinate-level scene categories (e.g., indoor or outdoor) and even more specific basic-level categories (e.g., desert or meadow) (Greene & Oliva, 2009; Torralba & Oliva, 2003). This created a problem in Experiment 1, because participants had to judge the match between a label and segment, while labels in incompatible trials were randomly drawn from all alternative classes. Thus, they often referred to images that strongly differed from the current image in their global properties, making incompatible trials very easy to solve, regardless of segment quality. Indeed, a separate analysis of incompatible trials indicated that performance was very high and did not differentiate between segment or image types. To make incompatible trials more informative, we conducted Experiment 2.

## 4 Experiment 2

Experiment 2 was identical to Experiment 1, with the only difference that all labels in incompatible trials stemmed from the same image type (e.g., desert images could be preceded by the label "wheat field", but not "lighthouse" or "office"). In this way, it was less feasible to simply work by exclusion due to a mismatch in global image properties. Conceptually, this makes Experiment 2 more similar to fine-grained classification (Lai et al., 2020; Rong et al., 2021). For this task, segments do not only have to be inspected more carefully, they also have to specifically target the most diagnostic image areas to benefit classification.

We hypothesised that these higher requirements would exacerbate the differences between our four segment types. Specifically, we expected the non-significant trends to become significant, making Grad-CAM worse than both human segment types overall, but clearly better than XRAI. However, we were most interested in whether the classification requirements would also affect the interaction between segment and image type. In Experiment 1, Grad-CAM had been descriptively worse than XRAI for object images in terms of error rates and we speculated that this was due to their suboptimal positioning (cf. Kapishnikov et al., 2019). We expected this positioning to become more consequential now, leading to slower and less accurate responses with Grad-CAM than all other segment types for objects. For landscapes and indoor scenes, we still expected performance to be worst with XRAI.

### 4.1 Method

#### 4.1.1 Participants

The general participant sample characteristics were identical to Experiment 1, as both experiments shared one and the same recruiting process: the first half of the people who registered were assigned to Experiment 1, and the second half to Experiment 2. Thus, 24 new participants took part in the experiment, 15 of them were female and 9 were male, and their age ranged from 18 to 38 years ($M$ = 24.9, $SD$ = 5.4). One additional participant took part, but her data were not analysed, because she was not a native German speaker and indicated problems in processing the labels quickly enough.

#### 4.1.2 Apparatus and stimuli

The apparatus and stimuli were identical to Experiment 1.

#### 4.1.3 Procedure

The procedure was identical to Experiment 1, with the exception that in incompatible trials, the label always corresponded to the other class of the same image type (e.g., when the segment was taken from an image of a lighthouse, the label was windmill, and vice versa). This led to the label-segment combinations presented in Table 1. Only these combinations were presented.

**Table 1**

*Possible combinations of labels and segments for incompatible trials in Experiment 2*

| Image type | Label | Segment |
|---|---|---|
| Objects | Lighthouse | Windmill |
| | Windmill | Lighthouse |
| Indoor scenes | Office | Dining room |
| | Dining room | Office |
| Landscapes | Desert | Wheat field |
| | Wheat field | Desert |

### 4.1.4 Data analysis

The data analysis procedures were identical to Experiment 1.

### 4.2 Results

### 4.2.1 Response times

The 4 (segment type) x 3 (image type) ANOVA revealed a main effect of segment type, $F(2.0,45.2) = 16.075$, $p < .001$, $\eta p^2 = .411$. This effect was due to participants responding more slowly to XRAI segments (1109 ms) than to gaze, drawing, and Grad-CAM segments (980, 924, and 1012 ms, respectively), all $ps < .010$, and more slowly to Grad-CAM than drawing segments, $p = .002$, while all other differences were non-significant, all $ps > .3$. Moreover, a main effect of image type, $F(2,46) = 35.943$, $p < .001$, $\eta p^2 = .610$, indicated that all image types differed from each other, all $ps < .001$, with responses being fastest for objects (934 ms), slowest for indoor scenes (1072 ms) and intermediate for landscapes (1013 ms). The main effects were qualified by a significant interaction, $F(6,138) = 23.198$, $p < .001$, $\eta p^2 = .502$, indicating that the effects of segment type strongly depended on image type (see Figure 5A). For objects, we replicated the finding that Grad-CAM segments (1059 ms) yielded slower responses than gaze and drawing segments (900 and 828 ms, respectively). However, in contrast to Experiment 1, this XAI cost was less clear for XRAI segments (947 ms), which enabled faster responding than Grad-CAM segments, $p = .010$, and similar performance to gaze segments, $p = .470$. It only was inferior to drawing segments, $p = .004$. The latter were particularly effective in Experiment 2 and even yielded faster responses than gaze segments, $p = .023$. A very different picture emerged for indoor scenes, where we replicated the finding that XRAI segments (1292 ms) led to slower responses than all other segment types, all $ps < .001$, while gaze, drawing, and Grad-CAM segments (1026, 995, and 976 ms, respectively) did not differ from each other, all $ps > .4$. The results were similar for landscapes, where responses were slower for XRAI segments (1088 ms) than gaze and drawing segments (1014 and 947 ms, respectively), both $ps < .044$. However, the difference to Grad-CAM segments (1002 ms) did not reach significance, $p = .099$. Again, gaze, drawing, and Grad-CAM segments did not differ from each other, all $ps > .3$.

**Figure 5**

*Response times (A) and error rates (B) for Experiment 2, depending on segment type and image type. Error bars represent standard errors of the mean.*

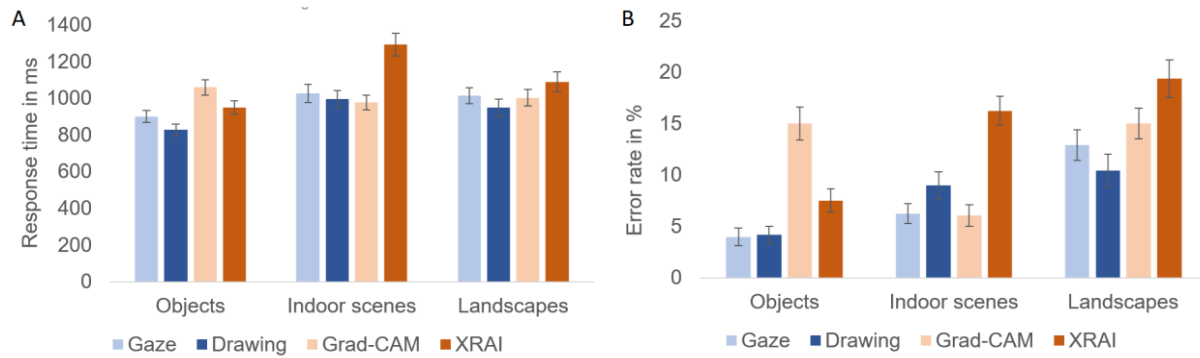

### 4.2.2 Error rates

In the error rates we found a main effect of segment type, $F(3,69) = 19.999$, $p < .001$, $\eta p^2 = .465$, indicating that overall, both human segment types yielded fewer errors than both XAI segment types. That is, error rates were lower for gaze and drawing segments (7.7 and 7.8 %, respectively) than for Grad-CAM and XRAI segments (12.0 and 14.4 %, respectively), all $ps < .009$. Neither the two human segment types nor the two XAI segment types differed from each other, both $ps > .3$. Second, a main effect of image type, $F(2,46) = 24.575$, $p < .001$, $\eta p^2 = .517$, resulted from more errors for landscapes (14.4 %) than objects and indoor scenes (7.7 and 9.4 %, respectively), both $ps < .001$, while error rates for objects and indoor scenes did not differ, $p = .152$. Third, there was a significant interaction, $F(6,138) = 10.613$, $p < .001$, $\eta p^2 = .316$, indicating that the differences between segment types depended on image type (see Figure 5B). For objects, more errors were committed with Grad-CAM segments (15.0 %) than all other segment types, all $ps < .004$, while no differences appeared between gaze and drawing segments (4.0 and 4.2 %, respectively), $p > .9$, but also XRAI segments (7.5 %) did not differ from the two human segment types, both $ps > .140$. Conversely, for indoor scenes, XRAI segments (16.2 %) yielded higher error rates than gaze, drawing, and Grad-CAM segments (6.2, 9.0, and 6.0 %, respectively), all $ps < .001$, while these three segment types did not differ from each other, all $ps > .3$. For landscapes, XRAI segments (19.4 %) led to higher error rates than gaze and drawing segments (12.9 and 10.4 %, respectively), while these two did not differ, $p > .6$, and Grad-CAM segments (15.0 %) did not differ from any of the other segment types, all $ps > .2$.

### 4.3 Discussion

Experiment 2 replicated the main findings of Experiment 1, but also extended them by more clearly differentiating between the two XAI methods. The main findings we replicated were that humans could more easily deal with human segments than XAI segments, while their problems with XAI depended on the specific XAI method and image type. With Grad-CAM segments, performance decrements were only evident for objects but not for indoor scenes and landscapes. The opposite pattern emerged for XRAI segments, which enabled high performance for objects but created problems for indoor scenes and landscapes. In contrast to Experiment 1, the increased classification requirements led to a clearer differentiation of performance effects between segment types and an even stronger dependence of segment interpretability on image type. Before comparing our human performance results to the performance of our DNN when classifying the same segments, we will report a final human experiment. This experiment examined whether the effects observed in objective performance measures could also be picked up by subjective ratings of segment quality.

# 5 Experiment 3

While conducting Experiment 1, it turned out that the experimental sessions took less time than anticipated. This gave us the opportunity to complement our objective performance assessment with a subjective rating of segment quality. That is, after participants had completed the classification task (Experiment 1 or 2), we asked them to evaluate each segment. To this end, they saw the correct class label next to a segment and had to indicate how well the segment reflected this label. Our hypotheses were analogous to Experiment 2. This is because we expected the ratings to mirror performance measures, while retaining the high specificity that can be achieved when contrasting segments with meaningful alternatives. In comparing subjective ratings and performance, we aimed to qualitatively analyse individual images to investigate not only the average differences but also the variations in each measure. However, this analysis was highly exploratory, thus we had no hypotheses about its results.

## 5.1 Method

### 5.1.1 Participants

The general participant characteristics were the same as in the previous two experiments, because Experiment 3 was performed as the second part of an experimental session for a subset of participants. Accordingly, 7 participants had previously taken part in Experiment 1 and 25 in Experiment 2. Of these 32 participants who joined Experiment 3, 19 were female and 13 were male, and their age ranged from 18 to 72 years ($M$ = 27.2, $SD$ = 11.2).

### 5.1.2 Apparatus and stimuli

The apparatus was identical to Experiments 1 and 2. Stimuli consisted of a segment on the left-hand side of the screen and a word plus rating scale on the right-hand side. Segments corresponded to the 240 images used in Experiments 1 and 2 (i.e., 60 images overlaid with the masks of four segment types). Words corresponded to the labels used in Experiments 1 and 2 and represented the six classes. However, in Experiment 3 they were presented on the same screen as the segments and were always compatible (i.e., participants always saw a segment with its correct label). Ratings were submitted on a ten-point scale with the question "How well does the segment depict the category?" and the anchors "very poorly" (1) and "very well" (10), presented in German. The scale consisted of ten numbered boxes with grey frames that changed their frame colour to yellow upon clicking them with the mouse, providing feedback that the click had been registered.

### 5.1.3 Procedure

All 240 segments were presented in random order. Note that participants already were familiar with these segments and their range of possible variation, as they had completed Experiment 1 or 2 just before. During Experiment 3, their task was to rate how adequately a segment depicted the class that was indicated by a label presented next to it. Participants submitted their rating by selecting the respective number box with their mouse and could alter their selection as often as they wanted. Once they were done, they had to press the Space bar to move on to the next trial.

### 5.1.4 Data analysis

The analysis procedure was the same as in Experiments 1 and 2, as ratings were subjected to the same 4 (segment type) x 3 (image type) ANOVA that had been used for response times and errors.

## 5.2 Results

### 5.2.1 Subjective ratings

The 4 (segment type) x 3 (image type) ANOVA revealed a main effect of segment type, $F(2.5, 76.4) = 287.638$, $p < .001$, $\eta_p^2 = .903$. This effect was due to participants giving higher ratings for gaze and drawing segments (7.9 and 7.8, respectively) than Grad-CAM and XRAI segments (7.0 and 6.3, respectively), all $p$s < .001. While Grad-CAM segments were also rated more favourably than XRAI segments, $p < .001$, the two human segment types did not differ, $p > .9$. A main effect of image type, $F(2, 62) = 88.811$, $p < .001$, $\eta_p^2 = .741$, further indicated that segments from object images (8.6) were rated higher than segments from indoor scenes and landscapes (6.4 and 6.7, respectively), both $p$s < .001, while the difference between the latter two did not pass the significance threshold, $p = .064$. Furthermore, there was a significant interaction, $F(3.9, 122.3) = 204.065$, $p < .001$, $\eta_p^2 = .868$, revealing that the effects of segment type varied with image type (see Figure 6). For objects, all differences between segment types were significant, all $p$s < .001. Although both human segment types yielded ratings close to the maximum value (10), ratings still were significantly higher for drawing than gaze segments (9.7 vs. 9.4, respectively). XRAI segments (8.6) were rated lower than the two human segment types but clearly higher than Grad-CAM segments (6.7). For indoor scenes, XRAI segments (4.5) were rated lower than gaze, drawing, and Grad-CAM segments (7.0, 6.8, and 7.2, respectively), all $p$s < .001. Moreover, Grad-CAM segments were rated even higher than drawing segments, $p < .001$, but not lower than gaze segments, $p = .064$. The human segment types did not differ significantly, $p = .179$. Finally, for landscapes, XRAI segments (5.7) were rated lower than gaze, drawing, and Grad-CAM segments (7.2, 7.0, and 7.0, respectively), all $p$s < .001, while no significant differences were found between these three segment types, all $p$s > .053.

**Figure 6**

*Ratings from Experiment 3, depending on segment type and image type. Results are presented (A) in their original coding (i.e., higher values reflect better ratings) and (B) in inverse coding (i.e., higher values reflect discrepancy from optimum) to facilitate visual comparison with the results of Experiments 1 and 2 presented in Figures 4 and 5. Error bars represent standard errors of the mean.*

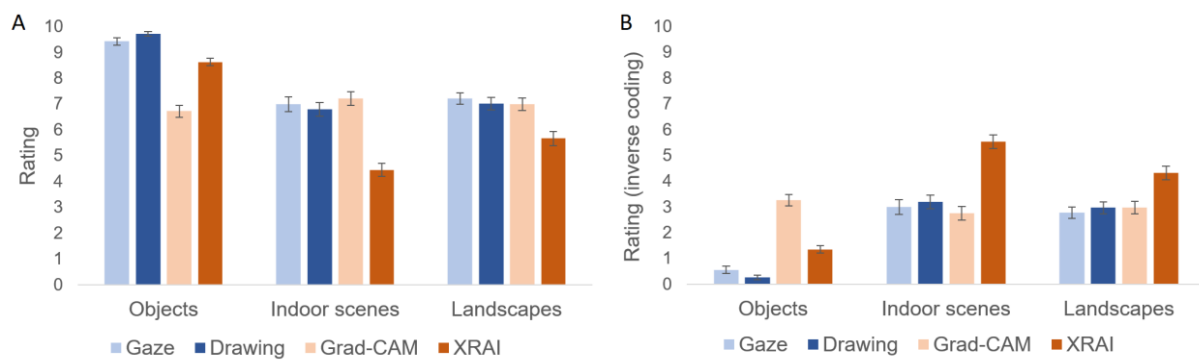

### 5.2.2 Qualitative assessment of image-specific effects

To understand how the effects of segment and image type depended on individual images, we looked at the inter-image variability in response times, error rates, and subjective ratings for each factor combination (see Figure 7). Overall, an inspection of the figure suggests that from left to right, the differentiation between design cells or factor combinations increased, while the differentiation within the cells decreased. That is, ratings seemed to produce more consistent values for a particular factor combination, with relatively little difference between individual images, at least for objects and indoor scenes. Response time data appeared somewhat more differentiated and error rates seemed most susceptible to image-specific effects. The figure also allows for some more specific observations. First, for objects, any indications of XRAI segments being inferior to human segments were driven by only two images, while the inferiority of Grad-CAM seemed rather consistent. The reverse was true for indoor scenes, where it was the inferiority of XRAI that seemed quite consistent. For deserts, the picture was somewhat less clear, as strong image-specific effects seemed to have occurred with all four segment types, perhaps because with these large-scale scenes it is less relevant which exact area is shown in a segment.

**Figure 7**

*Comparison of objective and subjective measures in their ability to reflect differences between image types and segment types. Each line represents an individual image. Darker shades of red indicate worse performance, and the coding of subjective ratings was inverted accordingly. Each cell contains the average values of 24 participants for errors and response times, and of 32 participants for ratings.*

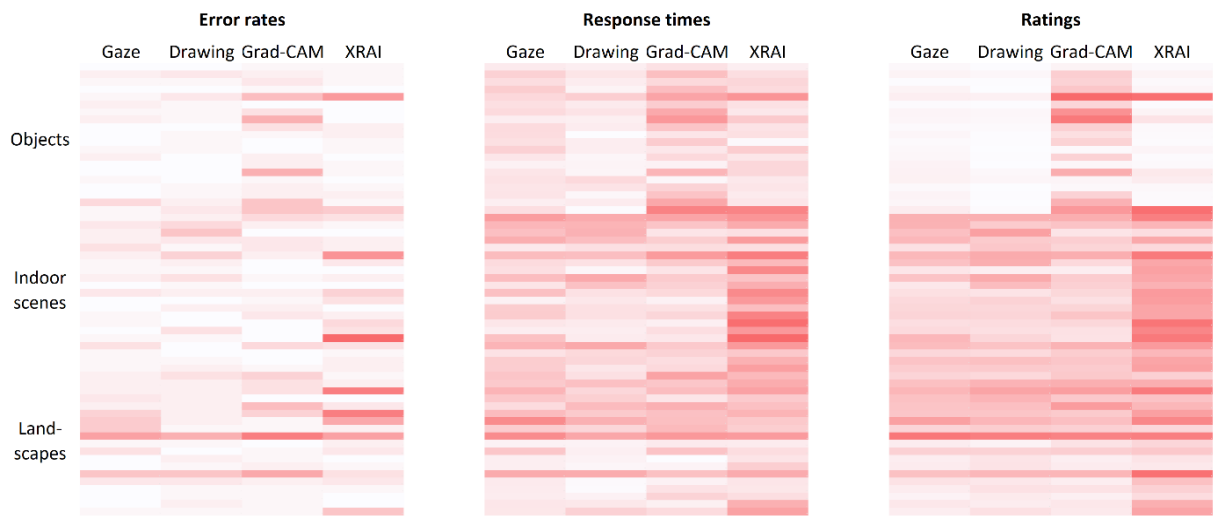

## 5.3 Discussion

The subjective ratings assessed in Experiment 3 generally mirrored the performance effects of Experiments 1 and 2, only with higher consistency and thus larger effect sizes. Once again, we found that on average, humans could deal better with human segments than XAI segments. However, again the type of XAI cost depended on image type, with Grad-CAM being rated less favourably only for objects and XRAI being rated less favourably only for indoor scenes and landscapes.

A possible methodological concern is that the participant sample of Experiment 3 included participants from both Experiments 1 and 2, although the two experiments imposed different classification requirements. In principle, the divergence in prior experience could have affected the ratings. Therefore, in a control analysis we only included participants from Experiment 2. However, the overall pattern of results was the same. Thus, the specificity requirements experienced before the rating assignment did not have any noteworthy impacts on participants' ratings. Given that our investigations of human interpretability yielded highly consistent results across three experiments, we can now compare these findings to the DNN's performance in classifying the same segments.

# 6 DNN performance

In the final step, we investigated whether our four segment types affected DNN classification in similar ways to humans. If humans and DNN benefit from the same visual information, DNN performance should mirror the results of the human experiments: inferior performance with Grad-CAM segments for objects, inferior performance with XRAI segments for indoor scenes and landscapes, and highest performance for human segments across all image types. However, we anticipated that the latter would not be realistic, because the XAI segments explained the very same DNN that now had to classify them. This is because the question we set out to answer was about the correspondence between fidelity and human interpretability: whether segments that were true to the DNN they were supposed to explain were also interpretable for humans. Given that the DNN now had to classify explanations of its own performance in the case of XAI segments, we expected higher performance with XAI segments than human segments overall. We had no specific hypotheses about whether the two types of human segments would differentially affect our DNN. On the one hand, several previous studies found that DNN benefitted from human eye movements, either when classified during testing (Liu et al., 2023; Rong et al., 2021) or when used for training (Boyd et al., 2023; Karargyris et al., 2021; Yang et al., 2022). However, we are not aware of any previous studies that contrasted DNN performance with different types of human attention maps in general or eye movements versus manual selection in particular.

We initially intended to test our DNN with only the 5 % area segments used in the human experiments. However, as we will see below, this only yielded acceptable performance under specific conditions. Therefore, we also tested how DNN performance would change with increasing segment size for the different segment types. Note that this comparison can only include human segments for rather small segment sizes, because large parts of several images received no human attention at all.

## 6.1 Method

To evaluate DNN performance on the segments, the original images were masked by changing all pixels to black (0, 0, 0) unless they were part of the segment. The resulting image was then fed into the network. We analysed true class certainty and top-5 accuracy. True class certainty denotes the likelihood that the DNN assigns to the correct class. This metric provides fractions of 1, with 1 being the total score summed over all 365 classes of the Places365 dataset. Thus, a value of 1 would result if the DNN decided that only the true class was a viable option, while all 364 alternative classes cannot possibly be correct. Top-5 accuracy can either be 1 or 0 for each image, indicating that the true class either is considered to be among the five most likely classes for this image or not. We then averaged the individual images' values across all 20 images included for a given combination of segment type and image type.

### Table 2

*Number of images included in each segment size for the human segment types. Numbers for the XAI segment types are not included as they were always 20 (i.e., all images were included).*

| | | Segment size in % of the total image area | | | | | | | | | |
|---|---|---|---|---|---|---|---|---|---|---|---|
| | | 5 | 10 | 15 | 20 | 25 | 30 | 35 | 40 | 45 | 50 |
| Objects | Gaze | 20 | 5 | 1 | 0 | 0 | 0 | 0 | 0 | 0 | 0 |
| | Drawing | 20 | 12 | 8 | 6 | 2 | 0 | 0 | 0 | 0 | 0 |
| Indoor scenes | Gaze | 20 | 20 | 19 | 10 | 5 | 0 | 0 | 0 | 0 | 0 |
| | Drawing | 20 | 20 | 20 | 20 | 20 | 20 | 20 | 20 | 20 | 19 |
| Landscapes | Gaze | 20 | 20 | 20 | 18 | 11 | 6 | 1 | 0 | 0 | 0 |
| | Drawing | 20 | 20 | 20 | 20 | 20 | 20 | 20 | 20 | 20 | 20 |

For the main comparison, we used the segments from the human experiments with a size of 5 %. However, as this hides much of the relevant image information, DNN performance was notoriously low, and thus we performed a second analysis in which we successively increased segment size in steps of 5 %, up to 50 %. Table 2 presents the numbers of images that could be included in this comparison for the human segment types. We were able to include all XAI segments at each step, as each image received at least some attention across its entire area.

## 6.2 Results

### 6.2.1 Segment size of 5 %

The DNN performance results for the 5 % segment size are summarised in Table 3. When averaged across the three image types, true class certainty was highest for drawing segments (.084), intermediate for XRAI segments (.057) and similarly low for gaze and Grad-CAM segments (.018 and .020, respectively). However, DNN performance was highly dependent on image type. Only objects yielded high performance (.123) and drove the overall results presented above: for this image type, the values were high for drawing segments (.238), medium for XRAI segments (.161), and low for gaze and Grad-CAM segments (.047 and .046, respectively). Conversely, for indoor scenes (.001), all segment types yielded consistently low performance (ranging from .000 to .002). For landscapes, the values were slightly higher (.010). Moreover, drawing and Grad-CAM segments yielded higher performance (.012 and .013, respectively) than gaze and XRAI segments (.007 and .008, respectively).

The top-5 accuracy results showed a somewhat different pattern. Overall, drawing segments (.317) still outperformed all other segment types. However, these three segment types yielded similar performance, although accuracy still was lowest for gaze segments, followed by Grad-CAM segments and then XRAI segments (.133, .150, and .167, respectively). In line with the true class certainty analysis, performance strongly depended on image type. Only objects yielded high performance (.488), and here the top-5 accuracy was highest for drawing segments (.800), medium for XRAI segments (.500) and similarly low for gaze and Grad-CAM segments (.350 and .300, respectively). Also in line with the previous analysis, it was impossible to classify indoor scenes with a 5 % segment size, yielding values of .000 for all segment types. Finally, for landscapes the values were low overall (.088) but somewhat higher for drawing and Grad-CAM segments (.150 for both) than for gaze segments (.050), while classification was impossible for XRAI segments (.000).

**Table 3**

*Mean values and standard deviations (in parentheses) for DNN performance in terms of true class certainty and top-5 accuracy, depending on segment type and image type*

|  |  | Objects | Indoor scenes | Landscapes |
|---|---|---|---|---|
| True class certainty | Gaze | .047 (.078) | .000 (.000) | .007 (.009) |
|  | Drawing | .238 (.241) | .001 (.001) | .012 (.014) |
|  | Grad-CAM | .046 (.067) | .001 (.001) | .013 (.016) |
|  | XRAI | .161 (.228) | .002 (.007) | .008 (.010) |
| Top-5 accuracy | Gaze | .350 (.489) | .000 (.000) | .050 (.224) |
|  | Drawing | .800 (.410) | .000 (.000) | .150 (.366) |
|  | Grad-CAM | .300 (.470) | .000 (.000) | .150 (.366) |
|  | XRAI | .500 (.513) | .000 (.000) | .000 (.000) |

*6.2.2 Stepwise increase of segment size*

The results of the stepwise increase in segment size are illustrated in Figure 8. In terms of performance changes with increasing segment size, a first notable observation was that classification performance did not improve but actually got worse for objects, while it improved considerably for indoor scenes and landscapes. Thus, larger areas were needed to correctly classify the scene-centric image types. The stepwise changes in DNN performance also revealed interesting interactions between segment type and image type. For objects (see Figure 8A and D), the segment types that originally enabled high performance at 5 % segment size (i.e., drawing, XRAI) showed a decrease in performance with increasing segment size. Such a decrease was also apparent for gaze segments but these results should be interpreted with caution, given that already at 10 %, three quarters of the images had dropped out of the set. For Grad-CAM segments, performance remained at a consistently low level. Taken together, the results for objects suggest that including other areas than the class-defining object tended to decrease DNN classification performance. The opposite pattern emerged for indoor scenes and landscapes, for which classification progressively got better with increasing segment size. The only segment type that deviated from this pattern was gaze, which initially increased, but only for top-5 accuracy, and then showed a performance drop (at least for landscapes) with larger segment sizes. However, it needs to be noted that for these larger segment sizes, only a fraction of the gaze segments remained in the set. For indoor scenes, XRAI segments yielded the highest performance overall, as well as the steepest increase. It surpassed all others segment types at about 25 % segment size, which is interesting given that XRAI segments had yielded worse performance than all others with 5 % segments in the human experiments. Grad-CAM segments were consistently harder to classify than XRAI segments. The results for drawing segments fell in between them. For landscapes, Grad-CAM segments were classified somewhat better than XRAI and drawing segments. However, overall the results and their increase with segment size were quite similar. Drawing segments seemed to reach a local plateau at around 30 %, and at around 35 or 40 % were surpassed by the two XAI segment types, which remained higher in accuracy than drawing segments for all larger segment sizes. Performance for gaze segments was inferior to all other segment types throughout.

**Figure 8**

*DNN performance change with increasing segment size in terms of true class certainty and top-5 accuracy, depending on image type and segment type. For the human segments, larger segment sizes only include the data of images in which attention was spread beyond the respective area size. Thus, incomplete lines indicate that no participant attended a larger area for any of the 20 images. The dot at a segment size of 100 % indicates DNN performance on the original, unmasked image.*

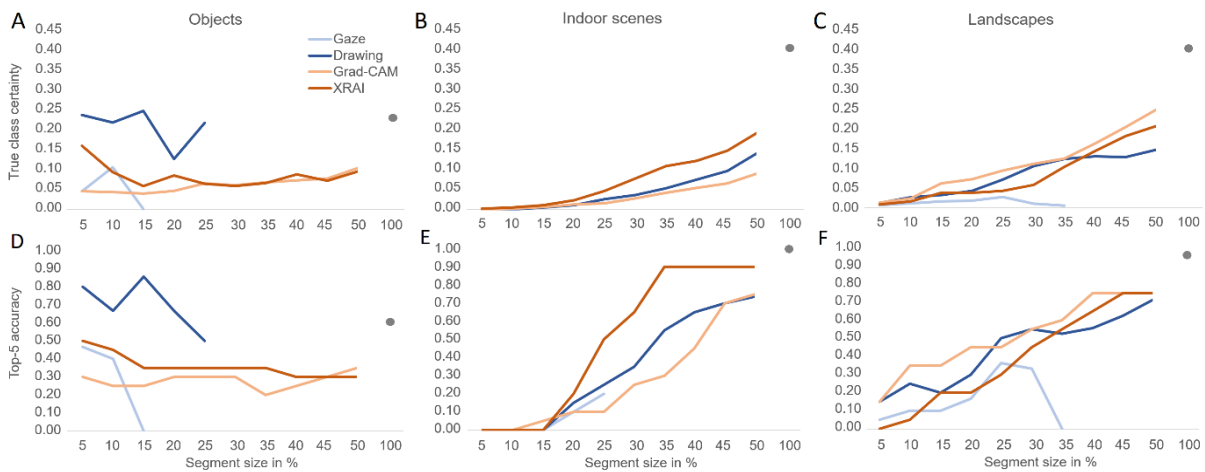

## 6.3 Discussion

We investigated whether the segment types that were most interpretable for humans also could be classified most accurately by the DNN that the segments were based on. In one way, DNN performance results replicated the human findings with regard to the differences between XAI segments. They again indicated that XRAI segments enabled a better classification of objects than Grad-CAM segments, while the reverse was true for landscapes. In another way, DNN performance clearly diverged from the human findings, namely with regard to the relative interpretability of the two human segment types. While both human segment types had yielded highly similar, good results in terms of human performance and ratings, drawing segments outperformed gaze segments by far in terms of DNN performance. In fact, drawing segments enabled higher DNN performance than any other segment type, including the XAI segments extracted from the very same network. This finding casts doubt on whether it is feasible to make general statements about the interpretability of human versus XAI attention maps, because it strongly depends on the way human attention is elicited.

## 7 General Discussion

A good XAI attribution method is one that can adequately explain an image classifier by highlighting relevant areas. But what does this mean? Three questions should be considered in this regard. First, do the attention maps reflect the information that is actually used by the DNN? Second, are they interpretable for humans? And third, do these two evaluation criteria go hand in hand? The contribution of this study is to specify how the notion of interpretability depends on evaluation context: the images and explanations that are used, and whether interpretability is assessed from a human or technical perspective. To elucidate the role of these context features, we presented the same image segments generated by two XAI methods (i.e., Grad-CAN, XRAI) to humans and DNN for classification. As a baseline for representing the information that humans actually need, we also included two types of human-generated segments (i.e., gaze, drawings). Human and DNN performance with these four segment types was compared for images that either focused on singular objects, object-to-object relations, or global scene properties (i.e., objects, indoor scenes, landscapes).

## 7.1 Overview of results

Across three experiments, humans had a harder time interpreting XAI segments than human segments overall, and more problems with XRAI than Grad-CAM segments. However, such general conclusions cannot be upheld as the effects were highly dependent on image type. Grad-CAM tended to generate suboptimal explanations for objects, with its segments often being ill-positioned and not targeting the most diagnostic parts of the object (cf. Kapishnikov et al., 2019; Rong et al., 2021). Conversely, Grad-CAM did quite well for indoor scenes and landscapes, where it often was on par with human segments. The reverse was true for XRAI. When looking at DNN performance, the differences between our two XAI segment types resembled the human findings: again, Grad-CAM led to inferior performance for objects, while XRAI was less interpretable for landscapes (indoor scenes could not be classified at all with the small segments). It is quite remarkable that the two XAI methods produced such different outcomes, given that they explained the same decisions of the same DNN.

How does the interpretability of XAI segments compare to that of human segments, and does it matter how the latter are generated? In this regard, humans and DNN differed quite fundamentally. For humans, the interpretability of the two human segment types was similar, except for a numerically small but significant drawing benefit for objects in response times and ratings. In contrast, for the DNN it had a tremendous impact which type of human segment needed to be classified. The DNN performed fairly well with drawing segments – even better than with XAI segments! This result is in line with a recent observation that the fidelity of human attention can be higher than that of common XAI methods (Liu et al., 2023). However, DNN performance plummeted with gaze segments. Given that the two human segment types provided similar contents (Müller, Dürschmidt, et al., 2023), this suggests that some aspects of the gaze visualisation were inconsequential for humans but highly problematic for DNN. We will return to this issue in the next section.

Taken together, while humans and DNN largely agreed on the interpretability of different XAI methods, they strongly disagreed on the interpretability of human explanations. Thus, we cannot provide a general answer to the question whether the same explanations are easily interpreted by humans and DNN. Instead, this depends on the procedure of eliciting explanations, the type of image, and the agent who needs to deal with the explanations. Thus, all these factors should be taken into account when evaluating XAI methods. Moreover, methodological decisions in the generation of explanations may affect the results, and two of them will be discussed below. Before turning to that, we first want to be transparent about the answers our study cannot provide.

### 7.2 Insights about the impacts of XAI: balancing applied versus basic research

There is an important question that we cannot answer with this study, namely to what extent XAI attention maps are useful. This is because we did not compare them to a baseline without explanations, and thus no general impacts of XAI (either benefits or costs) could be revealed. This distinguishes the present study from previous user studies that asked whether XAI attention maps supported human performance (Adebayo et al., 2022; Chu et al., 2020; Colin et al., 2022; Leichtmann et al., 2023; Nguyen et al., 2021; Shen & Huang, 2020). These studies provided XAI attention maps in addition to the original images. Conversely, we presented image segments. Thus, a baseline without XAI would necessarily have removed the mask that was hiding the remaining 95 % of the image. It is obvious that classification performance would have been higher for this baseline than for any segment.

From a practical perspective, one might therefore ask why we decided to use such a harmful visualisation of explanations – one that clearly makes the task harder for users, compared to a situation without explanations. In this regard, it is important to note that the present study is not an application study. Rather, we conducted basic research to investigate the usefulness of specific information contents selected by XAI. To assess this usefulness, one may have to deviate from the actual application context. This also becomes evident from previous studies, in which XAI consistently failed to improve human image classification performance, regardless of its quality (Chu et al., 2020; Nguyen et al., 2021). This makes sense when considering that humans can easily infer the meaning of a scene in an instant (Oliva, 2005). XAI attention maps might still support classification in some tasks, for instance when the class-defining area is hard to detect (Müller, Reindel, et al., 2023). However, they certainly are not needed to classify natural images in order to judge whether a DNN has made the right decision. If this classification is hard, for instance because the image is ambiguous or users do not know the class in principle, even faithful attention maps are unlikely to fix this problem (Colin et al., 2022; Leemann et al., 2023; Nguyen et al., 2021). Presumably, attention maps simply are not the right tool when the challenge is to answer questions about *what* is in an image, rather than *where* it is (Fel et al., 2023).

Regarding practical application, this means that one should critically consider which cognitive tasks can or cannot be supported by which types of XAI (Colin et al., 2022; Sanneman & Shah, 2022). Regarding basic research, it could mean that researchers may sometimes have to abstract away from the actual application context, for instance by presenting only the core information selected by a given XAI method. This still does not guarantee that significant effects of XAI will be found (e.g., Knapič et al., 2021). However, it will make it possible in principle to study non-trivial questions about the detailed cognitive mechanisms of using XAI attention maps. To this end, an interdisciplinary collaboration between computer scientists and cognitive psychologists seems like a fruitful approach.

### 7.3 Methodological issues of generating image segments

#### 7.3.1 DNN model and XAI methods

How appropriate are the technical methods we used in this study? It needs to be noted that our XAI segments were based on only two XAI methods and one particular DNN architecture and our artificial segment classification was also performed only by this one DNN. Thus, one might criticise that we do not provide evidence for the superiority of our technical methods over other state-of-the-art methods. Two things should be noted here. First, we did not propose any new DNN or XAI methods but used established ones to conduct a study on human-computer interaction. Accordingly, our aim was not for the DNN or XAI methods to be superior. Instead, we wanted them to be comparable to other methods used in similar studies, because this would increase the comparability of our results to related work. This is why we chose the DNN architecture (ResNet) and XAI method (Grad-CAM) that are by far the most common in this field. Still, we cannot make any claims about the generalisability of our results to other DNN and XAI methods. It is beyond the scope of this article to repeat our experiments with a large set of DNN architectures and XAI methods. However, our results suggest that the type of XAI method makes a large difference. We assume that the same will be true for the DNN architecture. It should be noted, that this actually is the main point we tried to make in this article: that it depends on the method. Future studies should specify what exactly this means. Specifically, how does the agreement between human and automated metrics for evaluating XAI vary for different DNN and XAI methods, used on different datasets, and supporting different tasks? Such comparative studies will provide a much more nuanced picture of what it means for XAI to be interpretable.

#### 7.3.2 Segment visualisation

A major point of disagreement between humans and DNN was their sensitivity to differences between the two human segment types: the DNN had considerable problems with gaze segments. At first glance, this seems to be at odds with two previous studies, in which human gaze was more conducive to DNN performance than XAI (Liu et al., 2023; Rong et al., 2021). However, the gaze elicitation procedure in that study was optimised for obtaining highly focused eye movements that only targeted a single diagnostic image feature. This leads to a possible explanation for our findings, namely that the DNN had problems with the fragmented gaze visualisation as a collection of blobs that were broadly spread across the image (at least for indoor scenes and landscapes, see Figure 1). Such fragmentation had not been an issue in previous studies, because the eye movements were either represented as segments but restricted to a single and highly localised image feature (Rong et al., 2021) or were represented as gradual heatmaps (Liu et al., 2023).

In our study, unlike the DNN, humans could easily ignore such low-level differences in visualisation and sustain a high level of performance as long as the relevant contents were visible. Presumably, this visibility was similar for both human segment types, as their areas had considerable overlap (Müller, Dürschmidt, et al., 2023). Remarkably, even subjective ratings hardly differed between the human segment types. One might conclude that humans paid little attention to visualisation details but were highly sensitive to changes in segment content. This also suggests a new perspective for future evaluations of XAI methods with human users, as these studies tend to be more concerned with visualisation than content (Karran et al., 2022; Sundararajan et al., 2019).

There are two complementary ways of testing whether the DNN's problems with gaze segments can actually be traced back to their visualisation. First, one could generate XAI segments that reflect the inner workings of the DNN, while mirroring the low-level visualisation details of gaze. This is not possible with the XAI methods we used, because they are intentionally designed to highlight connected regions. However, other XAI methods highlight individual pixels, such as Integrated Gradients (Sundararajan et al., 2017). Such methods could be adapted by using a visualisation procedure analogous to the one we used for our gaze segments (i.e., applying a Gaussian kernel with a diameter corresponding to human foveal vision, and multiplying it with DNN weights rather than fixation durations). Such gaze-like XAI segments could then be presented to the DNN. We do not claim that this is a suitable way of visualising XAI in practice. Rather, it would help us understand the relative impacts of deeper content features versus superficial presentation features on human versus DNN performance. Perhaps this would even lead us to revise our conclusion that humans and DNN largely agree on the interpretability of XAI. Perhaps, a more gaze-like XAI visualisation might leave human interpretability unaffected, but deteriorate the ability of DNN to make sense of the same segments. In this case, the interpretability of XAI outputs for humans and DNN would strongly diverge.

A second, complementary option would be to adjust the visualisation of gaze. Instead of using binary masks with sharp edges, attention maps could be visualised in a more gradual manner, comparable to the way XAI outputs are typically visualised (see Figure 2, second row). Occluding images with heatmaps would not make it possible to test how attention maps support classification. However, gradualness can alternatively be implemented by luminance, illuminating only those image areas that received sufficient attention (Giulivi et al., 2021; Leichtmann et al., 2023; Shen & Huang, 2020; Shitole et al., 2021). Such gradual visualisations might fundamentally change our results, because our DNN's problems with gaze should disappear if they were in fact due to the fragmentation and sharp edges in gaze segments. An additional benefit of gradual luminance maps is that they would make it transparent (quite literally!) how thoroughly an area was attended, which might provide additional guidance and facilitate classification.

### 7.3.3 Segment size

A second methodological aspect with major impacts on our results is the size of image segments. In line with previous work (Rong et al., 2021), DNN performance was low in general with segments of only five percent and only increased with larger segments. Moreover, the increases in segment size did not only have quantitative effects (i.e., more is always better) but also influenced the relative performance of the four segment types. In fact, it even reversed some of our conclusions, such as the inferior interpretability of XRAI segments for indoor scenes: with increasing segment size, the XAI method that had originally yielded the worst results now turned out to be the best one by far, reaching top-5 accuracies of .9 with image segments larger than 30 %.

In future research, it would be interesting to perform such stepwise uncovering procedures with humans as well (cf. Zhang et al., 2019) and compare the changes in performance to those of DNN. However, this would require images that have relatively large attention maps. For our eye movement recordings, this would not have worked out, given that many images did not contain even a single fixation outside the most important 5 % of their area. Even for indoor scenes and landscapes, the attended areas were relatively small (see Table 2). However, it would definitely be possible to find images on which human attention is spread more broadly. Given such images, it would be interesting to study under what conditions humans and DNN experience a similar increase in interpretability with an increase in visual evidence.

## 7.4 Limitations of the present study

A number of methodological and conceptual limitations might restrict the generalisation of our findings. For the human experiments, a limitation of the stimulus material is that we used a rather small set of images (60 in total, 20 per image type) and within image types we aimed for a high similarity. For instance, both lighthouse and windmill images presented a long, vertical building, with the most discriminative part being its top. Perhaps the problems Grad-CAM exhibited for objects would have been ameliorated if other objects had been used, for instance with all of their parts being similarly informative. The same goes for indoor scenes and landscapes. It is possible that indoor scenes with a higher diagnosticity of particular objects (e.g., a stove in a kitchen) would have yielded better results for XRAI when its segments contained only this one object. This would be in line with the finding that humans can easily infer scene classes from single objects if they occur frequently and specifically in the respective scene class (Wiesmann & Võ, 2023).

The recognisability of image segments goes hand in hand with a limitation in the way we implemented our basic task. As participants saw the label before the image, it is debatable whether their mental processes were sufficiently similar to free classification. For instance, the requirement to match a label and image might have led them to work by exclusion. This would be problematic if the latter required systematically different image information than a typical classification task. Therefore, an alternative task implementation would be to present the classification alternatives only after the image (van Dyck et al., 2021). While this would create other problems, it would be desirable to learn how different task procedures affect the results.

Another set of limitations concerns our experimental design or the way we assigned segment types to participants and blocks. First, participants saw segments from the same images four times, which opens the door for confounds with preview and practice effects. Given that we counterbalanced the order of segment type blocks, systematic influences on the main effect of segment type are unlikely. However, we cannot exclude systematic influences on the interaction of segment and image type. This is because practice effects presumably depend on segment similarity, which strongly varied with image type (Müller, Dürschmidt, et al., 2023). For objects, the four segments taken from one and the same image were much more similar than for indoor scenes and landscapes. That is, participants might have seen four almost identical views of the same lighthouse, but four rather different parts of the same office. A related concern results from the blockwise manipulation of segment type in Experiments 1 and 2, which might induce transfer effects between images. Such effects are conceivable in different directions, manifesting either as contrast effects or halo effects. An example for the latter is that participants might have perceived a given segment as more problematic after having seen suboptimal segments throughout the entire block. However, transfer effects are more likely to affect subjective ratings than response times and error rates. When we assessed these ratings in Experiment 3, the presentation was fully random and yet we observed the highest consistency. Nevertheless, future studies should test whether the present results can be replicated in other experimental designs such as between-subjects designs or designs with complete randomisation of all factors.

Concerning our DNN, a critical point is that we used the same network for segment generation and segment evaluation – in contrast to using different humans. This might be partly responsible for our finding that the DNN could better classify XAI segments than eye movements, which differs from the outcome of a previous study that trained a new DNN for segment evaluation (Rong et al., 2021). In our study, using the same DNN was necessary, given that we wanted to investigate how accurately the XAI segments reflected the DNN they were supposed to explain (and whether this went along with high human interpretability). However, one should keep in mind that we cannot conclude from our data whether humans and DNN benefit from the same image areas in general. Such inferences would require independence of the segment generation and evaluation processes.

Finally, a conceptual limitation arises from our basic approach of testing whether the interpretability of XAI segments for humans and DNN go hand in hand. We based our answer to this question on whether the same factors (i.e., segment type, image type) affect the two agents (i.e., humans, DNN) in the same way (e.g., both times revealing Grad-CAM problems with objects but XRAI problems with indoor scenes or landscapes). A different approach would be to generate a very large dataset of XAI segments that includes measures of human interpretability (e.g., reaction times, error rates, ratings) and measures of DNN interpretability (e.g., true class certainty, top-5 accuracy). These different kinds of measures could then be correlated statistically. In principle, such analyses would be possible within our design as well, but it would be hard to derive a meaningful interpretation of their results. This is because we used different image types and thus correlations could easily be misleading. For instance, a correlation could emerge from the mere fact that both human and DNN performance are high for objects and low for landscapes – even when within these image types human and DNN performance are not correlated at all. This is a well-known statistical phenomenon called Simpson's paradox (Simpson, 1951). Thus, performing separate analyses for each image type might be a better choice, but this would hardly make sense with the small number of images we presented for each image type (Schönbrodt & Perugini, 2013). In the future, large-scale studies should obtain human performance measures for large and varied datasets to assess the statistical relations between human and DNN performance.

### 7.4 Conclusion

Is it sufficient for evaluations of XAI methods to rely on automated fidelity metrics or should they also consider human interpretability? Our contribution to answering this question is an empirical user study to reveal under what conditions the two evaluation criteria go hand in hand or diverge. In contrast to previous studies, we varied three important influences: the image context, the general origin and specific method of explanations, and the agent interpreting them. In this way, we intended to provide a broad perspective on the evaluation of XAI quality. Indeed, our results underline the importance of taking all these factors into account. On the one hand, interpretability and fidelity do not seem to be completely incompatible concepts, because when XAI segments were easy to classify for humans under particular conditions, they also tended to be easy to classify for the DNN under the same conditions. However, automated XAI metrics do not seem sufficient, as we still found substantial differences between human and DNN performance. These differences were particularly striking when it came to human-generated segments. Relying only on automated metrics of these segments' explanation quality would have painted a completely different picture than relying on their usefulness for humans.

At the same time, the striking dependence on image type highlights a need to evaluate XAI methods on images with different characteristics, instead of merely relying on the common object-centric datasets that are often used in similar studies. Moreover, the particular visualisation of attention maps seems to play a large role for DNN but not for humans. Future research should explore the boundary conditions of these results, for instance by examining whether our conclusions can still be upheld when visualisations of XAI and human attention are more alike. Such differentiated comparisons would have practical value in that they can serve as a basis for better XAI evaluations. Moreover, they would help us understand the commonalities and differences in the processing mechanisms of humans and deep learning architectures.

## Acknowledgments

We want to thank Marcel Dürschmidt for support in generating the stimuli and Christopher Lindenberg for support in programming the DNN. Furthermore, we gratefully acknowledge the computing time provided by the center for high performance computing (ZIH) at TU Dresden.

## Funding

This work was supported by the German Research Foundation (DFG) under Grant PA 1232/15-1 and Grant TE 257/37-1.

## Declaration of interest statement

The authors report there are no competing interests to declare.

## Data availability statement

All images and human participant data are made available via the Open Science Framework (https://osf.io/pvmfj/). The source code for our DNN, XAI, and attention maps is made available on GitHub (https://github.com/cknoll/Humans-vs.-CNN-Effects-of-task-and-image-type).

## Author biographies

**Romy Müller** is a PostDoc at the Chair of Engineering Psychology and Applied Cognitive Research at TU Dresden. Her research focuses on the psychological mechanisms and domain-specificity of human-machine interaction in complex industrial systems. She is particularly interested in understanding and supporting human performance during fault diagnosis and decision-making.

**Marius Thoß** is a student in the Master program "Human Performance in Socio-Technical Systems" at TU Dresden. He is particularly interested in human-machine interaction and the role that explainable artificial intelligence can play in making complex technical systems more transparent and user-friendly.

**Julian Ullrich** is a student in the Master program "Computer Science" at the Heinrich-Heine Universität Düsseldorf. While working on Computer Vision and Reinforcement Learning problems, he developed a keen interest in explainable artificial intelligence in order to gain a deeper understanding of black box methods like DNN.

**Steffen Seitz** is a PhD student at the Chair of Fundamentals of Electrical Engineering at TU Dresden. His research interests revolve around neural network-based condition monitoring and explainable artificial intelligence. Specifically, he focuses on enhancing the understandability of model reasoning, aiming to make complex AI models more interpretable to human factory operators.

**Carsten Knoll** is a PostDoc at the Chair of Fundamentals of Electrical Engineering at TU Dresden. In his current research, he is mainly interested in explainable artificial intelligence for computer vision and in formal knowledge representation applied to control engineering and system dynamics.